\documentclass[10pt,letterpaper,twocolumn]{article}

\usepackage{ol2}
\usepackage{amsmath}
\usepackage{amssymb}
\usepackage[francais,english]{babel}
\usepackage{color}
\usepackage[T1]{fontenc}
\usepackage{graphicx}
\usepackage[draft]{hyperref}
\usepackage{psfrag}

\begin{document}

\twocolumn[
\title{Silicon-on-insulator integrated source of polarization-entangled photons}
\author{Laurent Olislager,$^{1,*}$ Jassem Safioui,$^{1,*}$ St\'ephane Clemmen,$^2$ Kien Phan~Huy,$^3$ Wim Bogaerts,$^4$ Roel Baets,$^4$ Philippe Emplit,$^1$ and Serge Massar$^5$}
\address{
$^1$OPERA--Photonique, CP~194/5, Universit\'e libre de Bruxelles (U.L.B.), av. F.D. Roosevelt 50, B-1050 Brussels, Belgium.\\
$^2$School of Applied and Engineering Physics, Cornell University, Ithaca, New York 14853, USA.\\
$^3$ D\'epartement d'Optique P.M. Duffieux, Institut FEMTO--ST, Unit\'e Mixte de Recherche du CNRS 6174, Universit\'e de Franche-Comt\'e, route de Gray 16, F-25030 Besan\c con, France.\\
$^4$Photonics Research Group, INTEC, Ghent University--IMEC, Sint-Pietersnieuwstraat 41, B-9000 Ghent, Belgium.\\
$^5$Laboratoire d'Information Quantique, CP~225, Universit\'e libre de Bruxelles (U.L.B.), av. F.D. Roosevelt 50, B-1050 Brussels, Belgium.\\
$^*$Corresponding authors: \href{mailto:lolislag@ulb.ac.be}{lolislag@ulb.ac.be} and \href{mailto:jsafioui@ulb.ac.be}{jsafioui@ulb.ac.be}
}
\begin{abstract}
We report the experimental generation of polarization-entangled photons at telecommunication wavelengths using spontaneous four-wave mixing in silicon-on-insulator wire waveguides. The key component is a 2D coupler that transforms path entanglement into polarization entanglement at the output of the device. Using quantum state tomography we find that the produced state has fidelity 88\% with a pure non-maximally entangled state. The produced state violates the CHSH Bell inequality by $S=2.37\pm0.19$.
\end{abstract} 
\ocis{130.0130, 270.0270, 270.5565.}
]

The ability to create, manipulate and transmit the quantum state of photons has enabled applications such as quantum key distribution, as well as foundational experiments concerning, for instance, quantum nonlocality and quantum teleportation. Polarization constitutes one of the degrees of freedom most often used to code quantum information. High-quality polarization-entangled photon-pair sources have been reported based on both $\chi^{(2)}$ (see e.g., \cite{1995-Kwiat, 2007-Fujii, 2010-Martin}) and $\chi^{(3)}$ (see e.g., \cite{2004-Takesue, 2005-Li}) nonlinear processes. To minimize cost and footprint, recent work focuses on the integration of such sources. The silicon-on-insulator (SOI) platform, based on reliable and low-cost CMOS technology, is a promising avenue for integrated photon-pair sources based on spontaneous four-wave mixing, both in straight wire waveguides \cite{2006-Sharping} and in ring resonators \cite{2009-Clemmen}. Earlier works reported time-bin entanglement \cite{2007-Takesue, 2008-Harada}, and polarization entanglement based either on a nonintegrated polarizing beam splitter (PBS) \cite{2008-Takesue} or on a polarization rotator sandwiched between two nonlinear silicon wire waveguides \cite{2012-Matsuda}. Here we present an SOI integrated source of polarization-entangled photons in which two nanophotonic waveguides produce path entanglement that is subsequently converted at the output of the chip into polarization entanglement using a 2D grating coupler. This is analogous to previous bulk or fiber optics experiments, such as \cite{2007-Fulconis}. We characterize the source using two-photon interferences, quantum state tomography, and Bell inequality violation.

\begin{figure}[b]
\centerline{\includegraphics[width=.75\columnwidth]{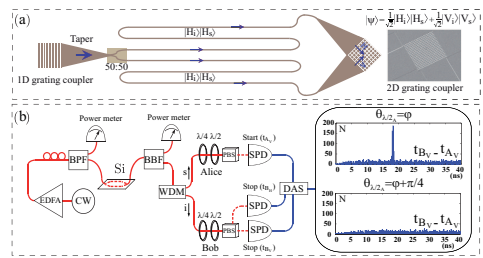}}
\caption{(a) Schematic of the SOI chip producing polarization-entangled photon pairs. Inset: SEM image of the 2D grating coupler. (b) Experimental setup for generating and measuring polarization-entangled photons (see text for detailed description of components). Inset: typical experimental results in the case of constructive and destructive interference.}
\label{fig:setup}
\end{figure}

Our SOI source [Fig.~1(a)] was fabricated by the ePIXfab at IMEC with 193\,nm deep UV lithography. A pump beam is coupled into the structure using a 1D grating coupler followed by a taper. A 50/50 multimode coupler \cite{2010-Bogaerts} then splits the light into two silicon wire waveguides. The waveguides have transverse dimension 500\,nm\,$\times$\,220\,nm and length 15\,mm. At the operating wavelength (telecommunication C band) the waveguides are monomode and guide only TE (horizontal) polarization. Four-wave mixing in the waveguides leads to photon-pair production, and hence to the state $a'\vert H_{s1}\rangle \vert H_{i1}\rangle + b'\vert H_{s2}\rangle \vert H_{i2}\rangle$, where $s,i$ refer to signal and idler frequencies, $1,2$ refer to the first and second waveguides, and we have indicated that the polarization is horizontal. The coefficients $a'$ and $b'$ take into account possible deviations from a perfect 50/50 coupler, or different losses in the two waveguides. The light propagating in the waveguides is coupled into an optical fiber using inverted tapers and a 2D grating coupler \cite{2003-Taillaert}, see inset of Fig.~1(a). Two-dimensional grating couplers enable polarization-insensitive SOI structures, as they couple the two orthogonal polarizations propagating in an optical fiber to the two TE modes propagating in two distinct silicon waveguides. They can provide an extinction ratio between both polarizations higher than 18\,dB \cite{2003-Taillaert}. In our case the 2D grating coupler converts path entanglement into polarization entanglement. The state in the optical fiber is thus $\vert \Psi(a,b)\rangle = a\vert H_s\rangle \vert H_i\rangle + b\vert V_s\rangle \vert V_i\rangle$, where the new coefficients $a,b$ take into account possible polarization-dependent losses of the 2D grating coupler.

The experimental setup is depicted in Fig.~1(b). A 1\,mW CW laser at 1539.6\,nm is amplified to 7\,mW with an erbium-doped fiber amplifier (EDFA) and then spectrally filtered by a bandpass filter (BPF). Injection and extraction losses are both approximately equal to 6\,dB, and losses in each arm of the structure are close to 3\,dB. The 1.75\,mW power on chip is divided in both device arms. The output containing the transmitted pump and entangled photons is collected by a cleaved nonpolarization maintaining standard telecommunication fiber (SMF) and passes through a band-block filter (BBF) (with an isolation of more than 110\,dB) to reject the pump and a wavelength division multiplexer (WDM) to separate signal ($s$) and idler ($i$) photons into two different fiber channels, which are sent to Alice's (A) and Bob's (B) stations. The signal and idler ports are centered at 1530 and 1550\,nm, respectively, with a bandwidth of 20\,nm. Alice and Bob locally and independently analyze their photons using a free-space analyzer consisting of a quarter-wave plate ($\lambda/4$), a half-wave plate ($\lambda/2$), and a PBS.  Only one output of Alice's PBS ---corresponding to the vertical polarization component--- is available, while both outputs of Bob's PBS are available. The photons emerging from the three outputs are directed to superconducting single-photon detectors (SPDs, Scontel, efficiency 5$\%$, dark-count rate 10\,Hz, time resolution 50\,ps). There is roughly 20\,m of SMF from the chip to the analyzers. Total losses from after ejection to the detectors are estimated at 3.6\,dB for Alice, and 5.5 and 6.8\,dB for Bob's two channels. Electronic signals from the SPDs are directed to a data acquisition system (DAS) consisting of a time-to-digital converter (Agilent Acqiris, time resolution 50\,ps) connected to a computer. The DAS registers the relative times $t_{\mathrm{B}_\mathrm{V}}-t_{\mathrm{A}_\mathrm{V}}$ and $t_{\mathrm{B}_\mathrm{H}}-t_{\mathrm{A}_\mathrm{V}}$ between A's and B's detections and outputs a histogram of these events. When correlated photons are present, a coincidence peak emerges from these time-resolved measurements. The illustrative histograms [inset of Fig.~1(b)] provide examples of constructive and destructive interference (note that each time bin in the histograms is 250\,ps long). Data acquisition and treatment are entirely automated.

To quantitatively analyze the results, we define the raw and net numbers of coincident events $N_\mathrm{raw}=\int_{t_i}^{t_f} N(t) \,\mathrm{d}t$, $N_\mathrm{net}=N_\mathrm{raw}-(t_f-t_i)\tau_\mathrm{acc}$, where $\tau_\mathrm{acc}=\int_{t_f}^{t_\mathrm{max}} N(t) \,\mathrm{d}t /(t_\mathrm{max}-t_f)$ is the rate of accidental coincidences, and $N(t)$ is the number of coincidences at time $t=t_{\mathrm{B}_\mathrm{V}}-t_{\mathrm{A}_\mathrm{V}}$ or $t_{\mathrm{B}_\mathrm{H}}-t_{\mathrm{A}_\mathrm{V}}$. The time window for the signal has size $t_f-t_i=0.8\,$ns. Outside this window, up to the maximum measurable delay $t_\mathrm{max}$, only noise is present. From these quantities we deduce that in the case of constructive interference, coincidences are measured at a rate $\approx0.4\,$Hz and the coincidence-to-accidental ratio is approximately equal to 8. This rather low value (which could be increased by using resonators or filters) is due to the CW operation (which increases the effect of the intrinsic noise of SOI waveguides \cite{2012-Clemmen}) and to the high losses from chip to detector. Dark counts are almost negligible due to the use of superconducting detectors. When all the fibers are carefully attached to guarantee stable injection and ejection and avoid polarization drift, no active power or polarization stabilization is required, and measurements are repeatable for several hours.

\begin{figure}[b]
\centerline{\includegraphics[scale=.31]{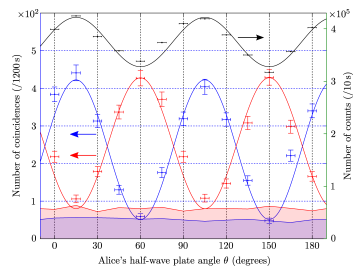}}
\caption{One- and two-photon interferences. Horizontal axis is the angle $\theta$ of Alice's half-wave plate ($\lambda/2_\mathrm{A}$), on which a precision of $\pm3^{\circ}$ is assumed. Left vertical axis is the number of coincidences between A's and B's detectors registered during 20\,min. Symbols are experimental results with statistical error bars, while curves are sinusoidal fits assuming a perfect net visibility. Shaded regions correspond to measured accidental coincidence rates. The blue and red curves (bottom curves) correspond to A$_\mathrm{V}$B$_\mathrm{V}$ and A$_\mathrm{V}$B$_\mathrm{H}$ coincidences, respectively. The black curve (top curve) is the single-photon rate of Alice's detector as a function of $\theta$. The right vertical axis is the total number of counts registered by Alice's detector in 10\,s.}
\label{fig:interf}
\end{figure}

Results of two-photon interference measurements are presented in Fig.~\ref{fig:interf}. Coincidence rates are plotted as a function of the angle of Alice's half-wave plate. Note that because no polarization management is realized before the analyzers, all phase plate angles must be adjusted to get a good contrast. Because of noise, raw visibilities are limited to approximately 80\% (vertical component) and 60\% (horizontal component), while net visibilities reach 99\% and 90\%, respectively. We also measured the single-photon rate detected by Alice [Fig.~\ref{fig:interf}, black (top) curve]. The fact that this curve is not perfectly flat is evidence that the produced state is not maximally entangled. This is presumably due to imperfect on-chip optical components. However, the limited visibility ($\approx12\%$) shows that the produced state is not far from a maximally entangled state. We note that nonmaximally entangled states have specific applications that are not accessible to maximally entangled states, see e.g., \cite{1993-Hardy, 2012-Giustina}.

\begin{figure}[t]
\centerline{\includegraphics[scale=.31]{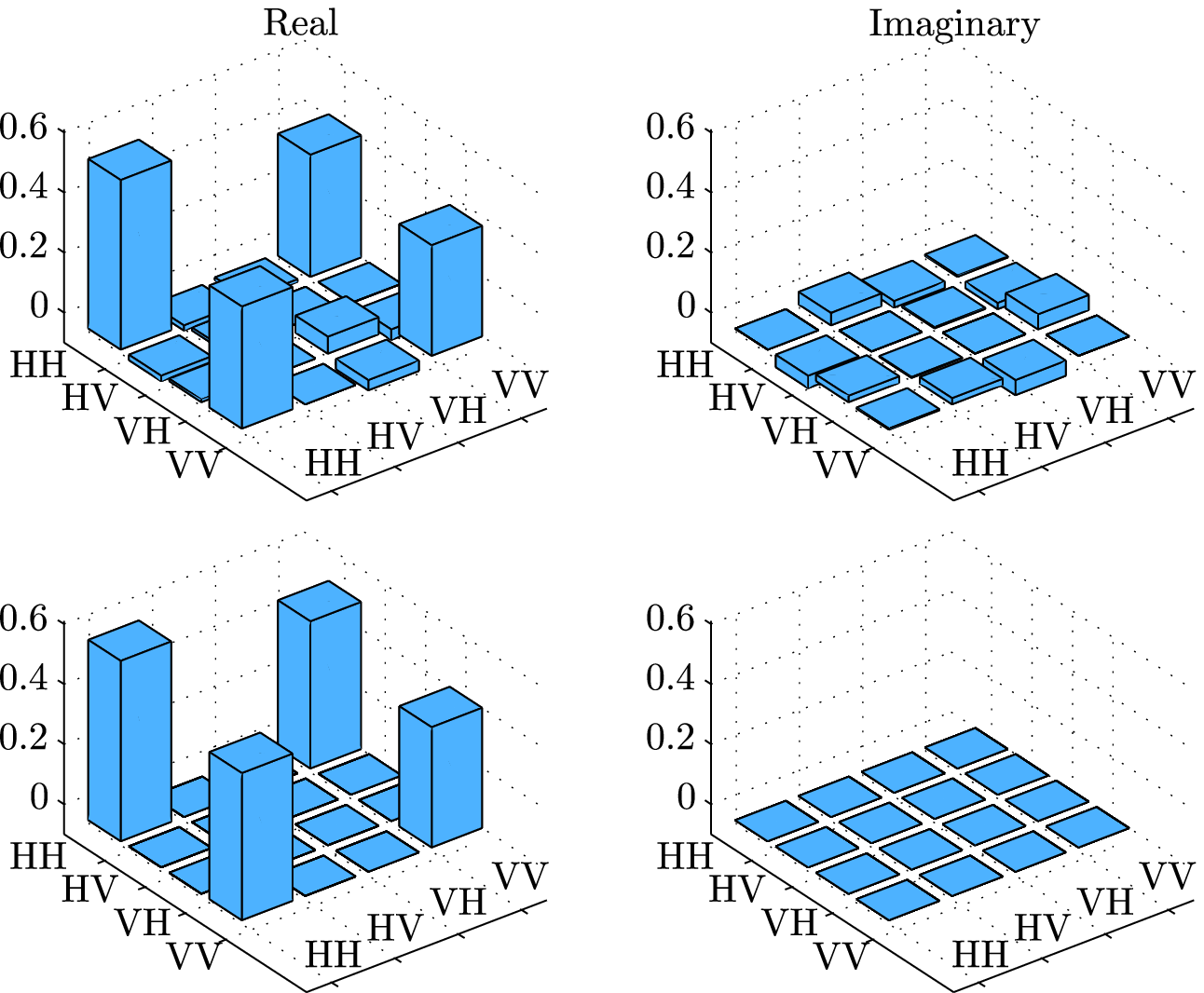}}
\caption{Estimated density matrix $\rho_\mathrm{AB}^{(\mathrm{in})}$ of the state produced at the output of the silicon chip is shown in the upper panels. This state has 88\% fidelity with the nonmaximally entangled state $\sqrt{0.6}\left|H\right\rangle_\mathrm{A} \left|H\right\rangle_\mathrm{B} + \sqrt{0.4}\left|V\right\rangle_\mathrm{A} \left|V\right\rangle_\mathrm{B}$, whose density matrix is represented in the lower panels. For each density matrix, $\mathrm{Re}(\rho)$ and $\mathrm{Im}(\rho)$ are plotted on the left and right, respectively.}
\label{fig:matrix}
\end{figure}

In order to accurately characterize the produced state we realized a standard quantum state tomography analysis. We used the measurements defined in \cite{2001-James} and followed the maximum likelihood method described therein to evaluate the "most probable" density matrix given the statistics of coincident events measured in 16 different configurations of the analyzers. The reconstructed density matrix $\rho_\mathrm{AB}^{(\mathrm{out})}$ is then re-expressed as 
$\rho_\mathrm{AB}^{(\mathrm{out})} = ( J_\mathrm{A} \otimes J_\mathrm{B} ) \rho_\mathrm{AB}^{(\mathrm{in})} ( J^\dagger_\mathrm{A} \otimes J^\dagger_\mathrm{B} )$. We optimized (numerically) the parameters of the Jones matrices $J_\mathrm{A} $ and $J_\mathrm{B}$ as well as the real numbers $a,b$ (with $a^2+b^2=1$) in order to maximize the fidelity $F\Bigl(\rho_\mathrm{AB}^{(\mathrm{in})},\rho_\mathrm{AB}^{(\mathrm{target})}\Bigr) = \Bigl( \mathrm{tr} \Bigl[ \Bigl( \sqrt{\rho_\mathrm{AB}^{(\mathrm{in})}} \, \rho_\mathrm{AB}^{(\mathrm{target})} \sqrt{\rho_\mathrm{AB}^{(\mathrm{in})}} \Bigr)^{1/2} \Bigr] \Bigr)^{2}$, where $\rho_\mathrm{AB}^{(\mathrm{target})} = \vert \Psi (a,b)\rangle\langle \Psi (a,b)\vert$ is the density matrix of the pure nonmaximally entangled state defined above. Thus, $\rho_\mathrm{AB}^{(\mathrm{in})}$ is our reconstruction of the state at the output of the SOI chip, and $J_\mathrm{A}$, $J_\mathrm{B}$ our reconstruction of the polarization rotation undergone by A and B photons between the chip and the analyzers. The results of this analysis are presented in Fig.~\ref{fig:matrix}. The reconstructed density matrix $\rho_\mathrm{AB}^{(\mathrm{in})}$ has a fidelity of 88$\%$ (which drops to 71$\%$ when noise is not subtracted) with the target state with $a^2\approx0.6$ and $b^2\approx0.4$. The fidelity to a maximally entangled state is 87\%.

Finally, we measured the CHSH inequality \cite{1969-Clauser} $S = E(A_1B_1) + E(A_1B_2) + E(A_2B_1) - E(A_2B_2) \leq 2$ with $E = ((N_{00}+N_{11})-(N_{01}+N_{10}))/((N_{00}+N_{11})+(N_{01}+N_{10}))$, $N_{ij}$ being the number of coincidences registered at Alice's output $i=0,1$ and Bob's output $j=0,1$. The three available outputs give directly the values of $N_{10}$ and $N_{11}$. To estimate $N_{00}$ and $N_{01}$, we proceed similarly to \cite{1974-Clauser}, using the expression $E = ((N_0^\mathrm{B}-2N_{10})-(N_1^\mathrm{B}-2N_{11}))/(N_0^\mathrm{B}+N_1^\mathrm{B})$, where $N_i^\mathrm{B}=N_{0i}+N_{1i}$, $i=0,1$, are estimated from two-photon interference measurements, see Fig.~\ref{fig:interf}. After carefully selecting the values of analyzer parameters for which the value of $S$ will be maximal, we measure (after subtraction of noise) $S=2.37\pm0.19$, thereby violating the CHSH inequality by almost two standard deviations.

In summary, we have presented an SOI integrated source of polarization-entangled photons based on a 2D grating coupler. In future work the degree of entanglement of the source could be tuned on chip by modifying the ratio of the integrated coupler. Our work confirms the relevance of SOI for integrated quantum optics.

This research was supported by the Interuniversity Attraction Poles program of the Belgian Science Policy Office, under Grant No. IAP P7-35, "photonics@be". We thank Nam Nguyen for experimental support.

\end{document}